\documentclass[useAMS,usenatbib]{mn2e}

\usepackage{amssymb}
\usepackage{amsmath}

\usepackage{epsfig}

\begin{document}


\title{Do Anomalous Narrow Line Quasars Cast Doubt on Virial Mass Estimation?}


\author[Charles L. Steinhardt and John D. Silverman]{Charles L. Steinhardt and John D. Silverman \\ IPMU, University of Tokyo, Kashiwanoha 5-1-5, Kashiwa-shi, Chiba, Japan}

\date{\today}

\maketitle



\begin{abstract}
Anomalous Narrow-Line Quasars (ANLs) are a population of quasars with narrow H$\beta$, and sometimes [O{\small III}] broader than $\sim 1000$ km/s, in total comprising $\sim 10-30$\% (most likely $\sim 25$\%) of Type I quasars at $0.2 < z < 0.8$.  We find that virial masses using the H$\beta$ and Mg{\small II} lines systematically differ for ANLs by an average of as much as 0.5 dex.  Because the broad H$\beta$ component width increases in ANLs but Mg{\small II} does not, we might suspect H$\beta$-based virial masses for ANLs are wrong but Mg{\small II} masses are correct.  If this is due to an outflow reaching the lower-ionization potential H$\beta$ line, C{\small IV} masses will be similarly flawed.  However, we cannot be certain of this explanation without followup work, and may be unable to identify which quasars are ANLs at $z > 0.8$.  Therefore, it is essential that ANLs be well-understood and well-modeled in order to allow the use of virial mass estimators on large optical spectroscopic catalogs, particularly at $z < 0.4$ and $z > 2.0$ where only one broad line is available for use in mass estimation.
\end{abstract}



\section{Introduction}

The quasar "broad-line region" (BLR) from which high-velocity gas produces correspondingly broad spectral emission lines, with typical FWHM in the 2000-20000 km/s range.  Prominent emission lines visible in optical spectra at $z > 0.5$ include, from ionization potential placing them nearest to the central black hole, C{\small IV}, a broad component of H$\beta$, and Mg{\small II}.  

For some individual active galactic nuclei, reverberation mapping \citep{Peterson2004,Bentz2009} has been able to confirm the locations of broad emission lines.  In a time series of spectra for the same object, an increase in the continuum luminosity is followed, often hundreds of days later, by a similar flare in C{\small IV} and then H$\beta$.  Assuming the flare propagates outward at the speed of light, the delay can be used to infer a radius to BLR spectral lines.  It is hoped that broad-line region velocities are predominantly virial.  For reverberation-mapped quasars, this approximation combined with Kepler's Laws then allows the use of C{\small IV} and H$\beta$ to infer a black hole mass.  

An empirical relationship between the reverberation radii and the continuum flux allows `virial mass estimators' \citep{McLure2002,McLure2004,Vestergaard2006,Wang2009,Onken2008,Risaliti2009,Rafiee2011} that can be used on just one spectrum rather than a time series.  Mass estimates have allowed a dramatic improvement in our understanding of distant quasars in several ways, including the tight $M - \sigma$ correlation between black hole mass and stellar dispersion in the host galactic bulge \citep{msigma1,msigma2}, analysis of the quasar mass distribution as a function of redshift \citep{Vestergaard2008}, and analysis of the quasar mass-luminosity plane \citep{Steinhardt2010a,Steinhardt2010b}.  The last two of these entirely depend upon virial mass estimates, while the others would require virial mass estimates if considered at higher redshift.

However, C{\small IV} in particular may also be substantially broadened by radiation pressure and quasar outflows \citep{Marconi2009}.  If so, this extra velocity would result in an overestimate of the mass of the central black hole.  It has recently been reported that $\sim 25$\% of quasars at $0.2 < z < 0.8$ have anomalous narrow lines, and that these ANL quasars are associated with broadened H$\beta$ but not broadened Mg{\small II}.  In this paper, we consider the implications of this broadening on virial mass estimation.

In \S~\ref{sec:sample}, we describe our line-fitting for the H$\beta$/[O{\small III}] complex and identification of ANLs.  We also examine the response of other emission lines to an increased narrow H$\beta$ width.  The resulting response of virial mass estimates in ANLs is examined in \S~\ref{sec:vme}.  Finally, the possible implications of these objects upon the validity of virial mass estimation are discussed in \S~\ref{sec:discussion}.

\section{Quasars with Broad H$\beta$ Lines}
\label{sec:sample}

We investigate the properties of the H$\beta$ and [O{\small III}] lines, as in \citet{Steinhardt2011a}, with a fitting prescription similar to \citet{Shen2008}.  First, the catalog redshift is assumed and the continuum is fit at rest wavelengths $[4435,4700]$ and $[5100,5535]$ \AA.  The continuum is best-fit as a power law combined with a Fe template from \citet{Bruhweiler2008}, where the Fe template has two free parameters: (1) an amplitude and (2) convolution with a Gaussian of variable width.  Once the continuum and iron lines are removed, H$\beta$ line is fit with two Gaussians and the [O{\small III}] doublet with a pair of Gaussians, such that the [O{\small III}] doublet is constrained to have the 3:1 amplitude ratio physically required by the fine structure transitions involved and the two [O{\small III}] widths are required to be identical to the width of the H$\beta$ narrow component.  

\begin{figure}
 \epsfxsize=3in\epsfbox{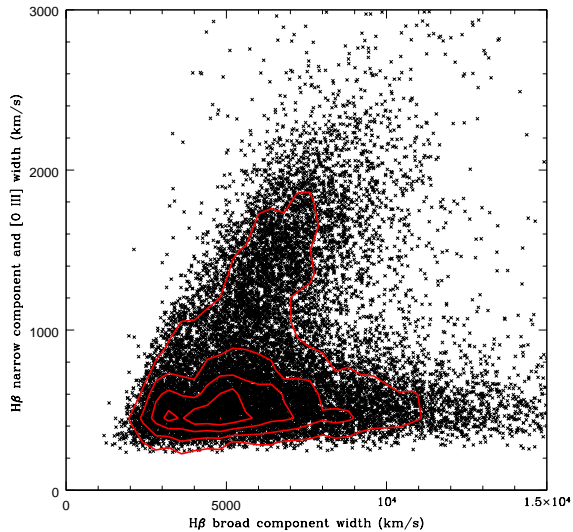}
\caption{Comparison of the [O{\small III}] FWHM and H$\beta$ broad-component FWHM for quasars in the SDSS DR7 catalog.  Lower narrow-line velocities are accompanied by a wide range of H$\beta$ widths, but anomalous narrow lines are well-correlated with broad H$\beta$, increasing with increasing broad H$\beta$ width.}
\label{fig:linecomp}
\end{figure}

A comparison between the H$\beta$ broad-line FWHM and combined narrow H$\beta$ and [O{\small III}] FWHM shows two populations (Fig. \ref{fig:linecomp}, a population with standard narrow lines uncorrelated with the H$\beta$ broad-line width and a population with broader ``narrow'' lines well correlated with the broad H$\beta$ component.  
The key question for virial mass estimation, then, is whether this broader H$\beta$ is due to an increased virial velocity or comes from non-virial broadening.  We divide the SDSS DR7 sample by narrow-line width, as in Table \ref{table:bins}.
\begin{table}
\begin{center}
\caption{Properties of quasars in the SDSS DR7 catalog binned by narrow-line width}
\begin{tabular}{|c|c|c|c|c|c|} 
\hline 
$\sigma$ (km/s) & FWHM (km/s) & Color & N & Fraction & $\log L_{bol}$ (erg/s) \\
\hline 
100--200 & 235--471 & Black & 3549 & 0.217 & 45.41 \\
200--300 & 471--706 & Black & 4941 & 0.303 & 45.51 \\
300--400 & 706--942 & Red & 2309 & 0.141 & 45.61 \\
400--500 & 942--1177 & Red & 1366 & 0.084 & 45.68 \\
500--600 & 1177--1413 & Yellow & 1086 & 0.067 & 45.72 \\
600--700 & 1413--1648 & Green & 1018 & 0.062 & 45.82 \\
700--800 & 1648--1884 & Cyan & 743 & 0.045 & 45.90 \\
800--900 & 1884--2119 & Blue & 424 & 0.026 & 46.00 \\
900--1000 & 2119--2355 & Magenta & 227 & 0.014 & 46.01 \\
\hline  
\end{tabular}
\end{center}
\label{table:bins}
\end{table}
We then co-add spectra within each bin to examine the dependence of Mg{\small II} emission on narrow H$\beta$ width.  Changes in other emission lines and the spectral continuum of ANLs are discussed in \citet{Steinhardt2011a} but are beyond the scope of this paper.

\begin{figure*}
 \epsfxsize=5in\epsfbox{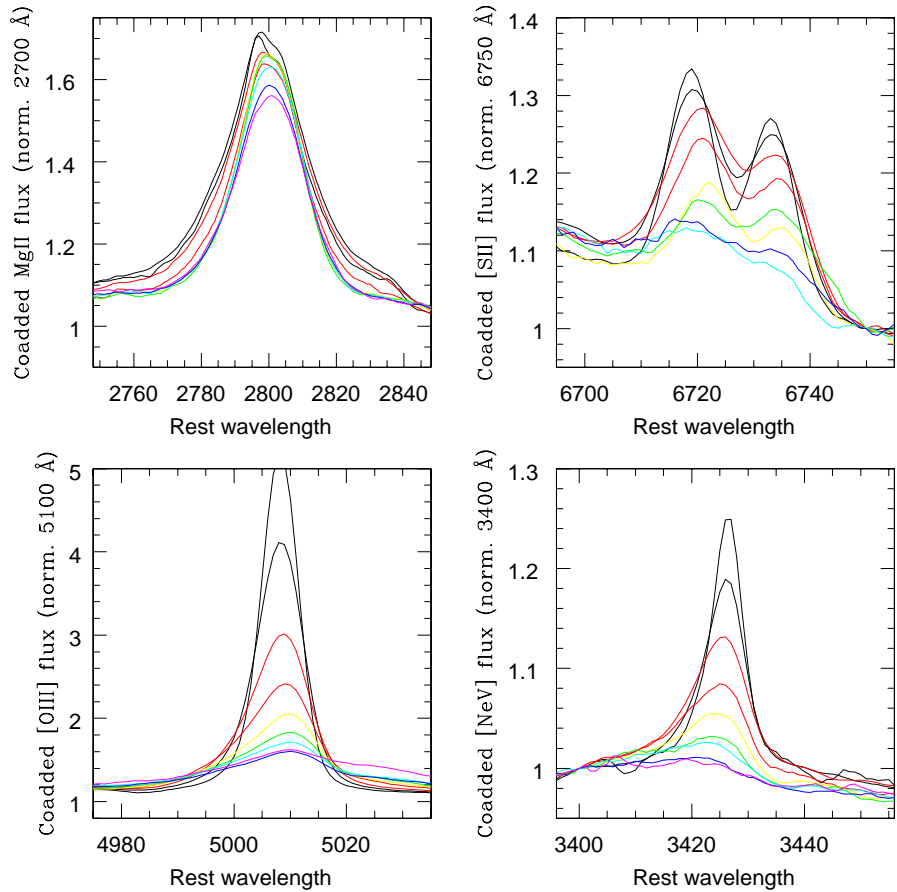}
\caption{Comparison of co-added quasar spectra near the Mg{\small II} (top left), [S{\small II}] (top right), [O{\small III}] (bottom left), and [Ne{\small V}] (bottom right) emission lines at different narrow-line widths binned and colored as in Table \ref{table:bins}.  At each wavelength, only spectra at redshifts where those wavelengths can be measured were co-added.  For [S{\small II}], there are too few objects at low $z$ to produce a magenta curve.}
\label{fig:lines}
\end{figure*}

Although the H$\beta$ FWHM increases in ANLs, the Mg{\small II} line width (Fig. \ref{fig:lines}) does not increase.  
Since Mg{\small II} has an ionization potential placing it at a larger radius from the central black hole \citep{Petersonbook}, one possible explanation might be an outflow propagating only partway through the broad-line region. 

It is also important to determine whether other narrow lines might be good indicators of ANLs, since [O{\small III}] is only visible at $0.2 \lesssim z \lesssim 0.8$.  For a the small fraction of our sample lying at $z < 0.39$, the [S{\small II}] $\lambda 6717,6730$ doublet is visible.  In standard quasars, [S{\small II}] is a pair of narrow lines, like [O{\small III}] and the narrow component of H$\beta$.  Although very few objects are included at high [O{\small III}] widths, it does appear that the [S{\small II}] amplitude may be decreasing with increased narrow-line width, while the [S{\small II}] width remains fixed (Fig. \ref{fig:lines}).  Still, [S{\small II}] appears to be a poor indicator of whether a quasar is an ANL.

However, [Ne{\small V}] $\lambda 3426$ may be a better indicator, as it broadens with broader ``narrow'' lines (Fig. \ref{fig:lines}).  As in \citet{Steinhardt2011a}, approximately 11\% of quasars are well-measured with broad [O{\small III}] and a broadened H$\beta$ narrow component, with an additional $\sim 15$\% showing anomalous narrow lines where H$\beta$ and [O{\small III}] are poorly matched, most commonly with more broadening found in H$\beta$ than in [O{\small III}] .  Thus, although [O{\small III}] and [Ne{\small V}] may turn out to be equally good indicators of ANLs, they are likely not good enough to identify the full set.  It may, however, be that finding these two forbidden lines broadened together, while non-forbidden lines and one other forbidden line are not, is useful in developing a physical model for ANLs.

\section{ANLs and Virial Mass Estimation}
\label{sec:vme}

One of the possible explanations for broader narrow H$\beta$ and [O{\small III}], correlated with a broad H$\beta$ component, is a strong outflow broadening all three lines.  Such an outflow would result in substantial non-virial motion in the H$\beta$ broad component.  Recent work using large catalogs to investigate the properties of quasar accretion has been greatly enhanced by virial mass indicators \citep{McLure2002,McLure2004,Vestergaard2006}, predicated amongst other assumptions upon the idea that the gas producing H$\beta$'s broad component is virialized.  This assumption has been questioned in more recent work \citep{Wang2009,Rafiee2011}, in which the virial velocity for broad lines is fit as $v^2 = \textrm{FWHM}^\gamma$ rather than $v^2 = \textrm{FWHM}^2$.  \citet{Rafiee2011} use various statistical methods to find for Mg{\small II} a series of best-fit values $1.21 < \gamma < 3.95$, arguing that the best of these techniques is the one yielding $\gamma = 1.21 \pm 0.40$, within 2$\sigma$ of $\gamma = 2$.  

A strong outflow would produce additional broadening of H$\beta$, and thus $\gamma < 2$.  Broader narrow lines are not, however, correlated with broadening in Mg{\small II} (Fig. \ref{fig:lines}).  Mg{\small II} and H$\beta$ masses show no offset, with a scatter of 0.22 dex \citep{Shen2008}.  Thus, if Mg{\small II} remains virial, a comparison between Mg{\small II}-based virial masses and H$\beta$-based virial masses is a potential test for the nature of H$\beta$ broadening.  Systematically larger H$\beta$ masses in ANL systems would then indicate that H$\beta$ is due to non-virial motion such as an outflow, while continued agreement would indicate that H$\beta$ remains virial, perhaps broadened because ANL systems produce the proper ionization potential for H$\beta$ at a different radius.

The \citet{Shen2011} catalog produces Mg{\small II}-based virial masses using the prescription of \citet{McLure2004} (along other estimators), which has been previously shown to agree with H$\beta$ masses to within a scatter of $\sim 0.22$ dex \citep{Shen2008}.  However, these catalogs restrict not just [O{\small III}] but also the narrow component of H$\beta$ to $\leq 1200$ km/s in FWHM.  ANLs include a broader H$\beta$ narrow component, so restricting the FWHM to 1200 km/s will also underestimate the H$\beta$ narrow component width.  

As a result, the line flux assumed by the fit to represent the H$\beta$ broad component would actually have been a combination of the broad component and the wings of the narrow component, resulting in a broad component fit that was artificially too narrow. An underestimate of the H$\beta$ FWHM will result in an underestimate in the virial mass $M_{BH} \propto \textrm{FWHM}^2$.  

\begin{figure}
 \epsfxsize=3in\epsfbox{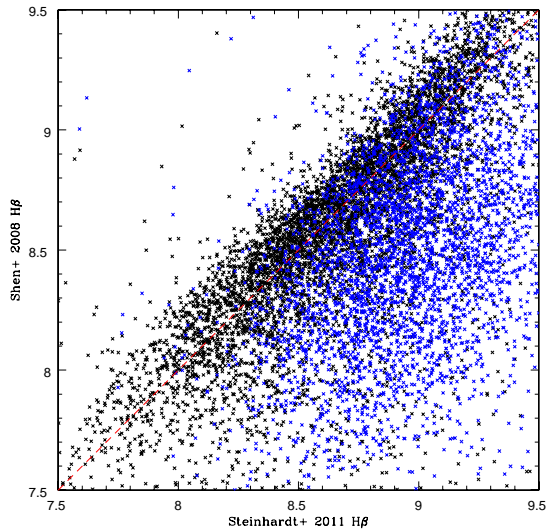}
\caption{Comparison of H$\beta$ masses between Shen et al. 2011, where the FWHM of [O{\small III}] and narrow H$\beta$ is constrained to be narrower than 1200 km/s, and masses produced in this paper, where these narrow lines are unconstrained.  The sample consists of 10133 well-measured quasars at $0.4 < z < 0.8$.  Black points are objects with [O{\small III}] narrower than 1200 km/s, while blue points correspond to objects with broader narrow components.}
\label{fig:hbcomp}
\end{figure}

Using the fitting routine described in \S~\ref{sec:sample}, we produce an alternative set of masses from the broad H$\beta$ component, where the narrow component is no longer constrained to lie at a FWHM less than 1200 km/s.  For narrow [O{\small III}] lines, there is good agreement between both sets of H$\beta$ masses, but for ANLs, the \citet{Shen2011} masses are systematically lower (Fig. \ref{fig:hbcomp}).  This is likely due to the effect described above, where a \citet{Shen2011} fit a combination of the broad and narrow component as their broad component.  Objects are only included in this sample if they have high enough signal-to-noise for our routine to produce a high-quality fit.  Note that the high signal-to-noise sample contains a larger ANL fraction than the overall sample, since ANLs are on average more luminous than other quasars.

For 4716 high signal-to-noise ANLs at $0.4 < z < 0.8$ (where both H$\beta$ and Mg{\small II} masses are available), our H$\beta$ masses average $0.474$ dex higher than \citet{Shen2011} using H$\beta$, while for 5417 other quasars our masses average $0.075$ dex higher.  Some of this discrepancy is due to changes in the Fe{\small II} template used and other line-fitting details, which are particularly important for Mg{\small II}.  Compared to the \citet{Shen2011} Mg{\small II} masses, our masses average $0.455$ dex higher for ANLs and $0.170$ dex higher for other quasars. 
\begin{table}
\begin{center}
\caption{Mean virial mass comparison between this work and Shen11 for well-measured quasars binned by [O{\small III}] width, $0.5 < z < 0.6$}
\begin{tabular}{|c|c|c|c|}
\hline 
FWHM (km/s) & $\log M/M_\odot$  & $\log M/M_\odot$ & $\log M/M_\odot$  \\
 & (H$\beta$, SS11) &  (Mg{\small II}, Shen11) &  (H$\beta$, Shen11) \\
\hline 
235--471 & 8.52 & 8.35 & 8.49 \\
471--706 & 8.63 & 8.48 & 8.61 \\
706--942 & 8.63 & 8.46 & 8.53 \\ 
942--1177 & 8.61 & 8.44 & 8.44 \\
1177--1413 & 8.67 & 8.37 & 8.30 \\
1413--1648 & 8.75 & 8.40 & 8.32 \\
1648--1884 &  8.83 & 8.44 & 8.35 \\
1884--2119 & 8.92 & 8.51 & 8.40 \\ 
2119--2355 & 9.02 & 8.53 & 8.46 \\
\hline  
\end{tabular}
\end{center}
\label{table:ourmasses}
\end{table}
Since the mass and luminosity distributions change with redshift \citep{Steinhardt2010a}, it is likely best to compare these mass estimates in a relatively narrow redshift range, yet one wide enough to include a large enough sample to draw meaningful conclusions.  At $0.5 < z < 0.6$ (Table \ref{table:ourmasses}), our H$\beta$ masses \citep{Vestergaard2006} average $0.047$ dex higher than \citep{Shen2011} for 1161 ``standard'' quasars (FWHM $< 1000$ km/s), but $0.393$ dex higher for 661 ANLs ($> 1000$ km/s).  More generally, the discrepancy between our unconstrained H$\beta$ masses and the \citet{Shen2008} Mg{\small II} masses, which are based upon a line that does not change in ANLs compared to other objects, grows with increasing [O{\small III}] width (Table \ref{table:ourmasses}).

For comparison, H$\beta$ and Mg{\small II} virial masses in the \citet{Shen2008} catalog show good agreement for all quasars.  If the uncertainty in comparing H$\beta$ and Mg{\small II} masses is uncorrelated between different quasars with a scatter of $\sim 0.22$ dex \citep{Shen2008}, the systematic error produced by restricting the narrow H$\beta$ component to 1200 km/s would be the dominant source of error in ANL masses.

Alternative estimates for the statistical uncertainty in H$\beta$ alone include 0.4 dex \citep{Vestergaard2006} and 0.21 dex \citep{Steinhardt2010c}, with the latter in addition to 0.15 dex scatter in Mg{\small II}.  Neither of these estimates investigated whether errors in H$\beta$ and Mg{\small II} masses are correlated, as they will be for errors emanating from the empirical relationship between monochromatic continuum luminosity and inferred radius to the broad-line region.  Even if the uncertainty were 0.4 dex and entirely statistical, the discrepancy between H$\beta$ and Mg{\small II} mass estimates for ANLs would be at least comparable.  Thus, this systematic disagreement cannot be neglected for individual objects, let alone in the analysis of large catalogs where the sample size allows a substantial reduction in the statistical error.

We also note that the \citet{Shen2011} masses show much better agreement between H$\beta$ Mg{\small II} for ANLs, despite requiring as part of the fitting routine that the narrow component of H$\beta$ along with the [O{\small III}] doublet have FWHM $\leq 1200$ km/s.  As above, an artificially-low broad H$\beta$ width will produce an artificially low virial mass $M_{\textrm{vir}} \propto \textrm{FWHM}^2$.  Remarkably, this lower H$\beta$ mass turns out to be empirically almost exactly the amount needed to bring H$\beta$ and Mg{\small II} back into agreement: \citet{Shen2011} H$\beta$ masses average $0.094$ dex higher for 5417 well-measured ``standard'' quasars (FWHM $< 1000$ km/s), but $0.075$ dex lower for 4716 well-measured ANLs ($> 1000$ km/s).  Similarly, at $0.5 < z < 0.6$, \citet{Shen2011} H$\beta$ masses average $0.111$ dex higher for 1161 ``standard'' quasars (FWHM $< 1000$ km/s), but $0.065$ dex lower for 661 ANLs ($> 1000$ km/s).  It should be noted that although H$\beta$ and Mg{\small II} masses appear close to agreement for both standard and ANL quasars, the typical offset has changed by $\sim 0.17$ dex between the two samples.

Regardless, these are not based upon the full FWHM of the H$\beta$ line.  As demonstrated by \citet{Rafiee2011}, without a theoretical basis for choosing a mass estimator, different empirical calibration techniques will produce a wide variety of possible masses due to the very small number of reverberation mapping systems available.  There is a theoretical basis for using the entire line FWHM and assuming virial motion in order to determine the virial component of the broad-line region velocity as $v_{\textrm{vir}} \propto \textrm{FWHM}^2$.  However, this relation now seems difficult to reconcile with the existence of ANLs even for H$\beta$, let alone C{\small IV} at a smaller radius from the central black hole.  It may be possible to develop a theoretical basis for reducing the measured FWHM to a purely virial FWHM if the broadening can be well-modeled.  However, we currently lack that understanding, and there is no basis for using the \citet{Shen2011} prescription for reducing the FWHM over any other that could be produced from calibrating H$\beta$ masses against Mg{\small II} for ANLs.

Virial mass estimation yields systematically different results for ANLs when using H$\beta$ and Mg{\small II}, even though the effect was previously masked by constraints in the \citet{Shen2008} line fitting routine.  At least one of the two must be wrong, either because of substantial non-virial motion in the broad line or because the empirical relationship between the radius to the broad-line region and continuum luminosity breaks down for ANLs.  Because the Mg{\small II} line does not appear to change with increasing [O{\small III}] while the broad component of H$\beta$ is strongly correlated, we might suspect that the H$\beta$ masses are in error, perhaps due to a strong outflow.  However, it is proper to be cautious regarding Mg{\small II} masses as well until there are better measurements, ideally including reverberation mapping, on ANLs and these calibrations can be examined more closely.

\section{Discussion}
\label{sec:discussion}

Virial mass estimates for anomalous narrow-line quasars (ANLs), a new population comprising perhaps one quarter of Type I quasars at $0.2 < z < 0.8$, appear to be flawed.  For ANLs, the assumption that the line FWHM is entirely due to the virial velocity of the gas producing H$\beta$ and Mg{\small II} leads to a systematic disagreement in estimated mass, with the disagreement intensifying with broader [O{\small III}].  We must conclude that at least one of these estimates is wrong.  

The prognosis for virial mass estimation is potentially dire.  The best-case scenario is likely that H$\beta$ is indeed broadened by an outflow, and we might still hope that Mg{\small II}, which appears unresponsive to that outflow, remains virial and can be used to produce a reliable mass.  Note, however, that ANLs represent $\sim 30$\% of quasars at $0.2 < z < 0.8$.  Therefore, at $0.2 < z < 0.4$, where Mg{\small II} does not lie within the SDSS spectrograph, 30\% of quasars will have an incorrect mass.  At present, that mass will be uncorrectable, although a sufficiently precise model for the outflow may allow a future correction.  

It is at least fortunate that the presence of [O{\small III}] (and the narrow H$\beta$ component) around the H$\beta$ line provides an indicator for many (but not all) of the 30\% of objects that will have incorrect masses.  However, much of the utility of virial masses has come from our ability to take a large, statistically-complete sample (such as provided by SDSS at $i < 19.2$) and look at the mass distribution, often compared against other quasar properties.  ANLs are not randomly-selected quasars, but rather quasars with specific emission line properties, continuum properties, and even host galactic properties.  Even removing all ANLs from the sample and only using the remaining 70\% of quasars will likely result in incorrect conclusions.

At high redshift, the problem becomes even worse.  Although the BOSS \citep{BOSS} survey will provide a large, statistically-complete sample at high redshift, for $z > 2.0$, Mg{\small II} is no longer visible and C{\small IV} masses must be used.  C{\small IV} has an ionization potential placing it even closer to the central black hole than H$\beta$ and is more vulnerable to effects such as radiation pressure and quasar wind \citep{Marconi2009}, and thus any outflow affecting H$\beta$ very likely will also affect C{\small IV}.  So, we might expect $\geq 30$\% of C{\small IV} masses to also be wrong due to outflows.  Yet, for H$\beta$, the presence of the [O{\small III}] narrow line at least allows a detection of ANLs and confidence that non-ANLs will have better-measured masses.  Although other narrow lines exist in quasar spectra, [S{\small II}] is not broadened in ANLs, even though [Ne{\small V}] is.  It is plausible that all narrow lines visible in the optical at $z > 2.0$ will remain narrow, with the strong semi-forbidden line C{\small III}] $\lambda 1909$ perhaps our best hope for an ANL indicator.  If so, not only will $\geq 30$\% of C{\small IV} masses be overestimates, but we may also have no indication as to which masses are erroneous.  Understanding ANLs well enough to model and outflows and correct for them would then become a requirement in order to make use of high-redshift quasar masses.

The authors would like to thank Steve Balbus, Tim Brandt, Forrest Collman, Martin Elvis, Eilat Glikman, Jeremy Goodman, Julian Krolik, Greg Novak, Jerry Ostriker, David Spergel, Michael Strauss, and Todd Thompson for valuable comments.  This work was supported by World Premier International Research Center Initiative (WPI Initiative), MEXT, Japan.

\bibliographystyle{apj}
\bibliography{ms}

\begin{thebibliography}{22}
\expandafter\ifx\csname natexlab\endcsname\relax\def\natexlab#1{#1}\fi

\bibitem[{Bentz} et~al.(2009){Bentz}, {Peterson}, {Netzer}, {Pogge} \&
  {Vestergaard}]{Bentz2009}
{Bentz} M.~C., {Peterson} B.~M., {Netzer} H., {Pogge} R.~W., {Vestergaard} M.,
  2009, \apj, 697, 160

\bibitem[{Bruhweiler} \& {Verner}(2008)]{Bruhweiler2008}
{Bruhweiler} F., {Verner} E., 2008, \apj, 675, 83

\bibitem[{Ferrarese} \& {Merritt}(2000)]{msigma1}
{Ferrarese} L., {Merritt} D., 2000, \apjl, 539, L9

\bibitem[{Gebhardt} et~al.(2000){Gebhardt}, {Kormendy}, {Ho} et~al.]{msigma2}
{Gebhardt} K., {Kormendy} J., {Ho} L.~C., et~al., 2000, \apjl, 543, L5

\bibitem[{Marconi} et~al.(2009){Marconi}, {Axon}, {Maiolino}
  et~al.]{Marconi2009}
{Marconi} A., {Axon} D.~J., {Maiolino} R., et~al., 2009, \apjl, 698, L103

\bibitem[{McLure} \& {Dunlop}(2004)]{McLure2004}
{McLure} R.~J., {Dunlop} J.~S., 2004, \mnras, 352, 1390

\bibitem[{McLure} \& {Jarvis}(2002)]{McLure2002}
{McLure} R.~J., {Jarvis} M.~J., 2002, \mnras, 337, 109

\bibitem[{Onken} \& {Kollmeier}(2008)]{Onken2008}
{Onken} C.~A., {Kollmeier} J.~A., 2008, \apjl, 689, L13

\bibitem[{Peterson}(2008)]{Petersonbook}
{Peterson} B., 2008, {An Introduction to Active Galactic Nuclei}, {Cambridge
  University Press: Cambridge}

\bibitem[{Peterson} et~al.(2004){Peterson}, {Ferrarese}, {Gilbert}
  et~al.]{Peterson2004}
{Peterson} B.~M., {Ferrarese} L., {Gilbert} K.~M., et~al., 2004, \apj, 613, 682

\bibitem[{Rafiee} \& {Hall}(2010)]{Rafiee2011}
{Rafiee} A., {Hall} P.~B., 2010, ArXiv:1011.1268

\bibitem[{Risaliti} et~al.(2009){Risaliti}, {Young} \& {Elvis}]{Risaliti2009}
{Risaliti} G., {Young} M., {Elvis} M., 2009, \apjl, 700, L6

\bibitem[{Ross} et~al.(2010){Ross}, {Sheldon}, {Myers} et~al.]{BOSS}
{Ross} N., {Sheldon} E.~S., {Myers} A.~D., et~al., 2010, in { Bulletin of the
  American Astronomical Society\/}, vol.~41 of { Bulletin of the American
  Astronomical Society\/},  517

\bibitem[{Shen} et~al.(2008){Shen}, {Greene}, {Strauss}, {Richards} \&
  {Schneider}]{Shen2008}
{Shen} Y., {Greene} J.~E., {Strauss} M.~A., {Richards} G.~T., {Schneider}
  D.~P., 2008, \apj, 680, 169

\bibitem[{Shen} et~al.(2011){Shen}, {Richards}, {Strauss} et~al.]{Shen2011}
{Shen} Y., {Richards} G.~T., {Strauss} M.~A., et~al., 2011, \apjs, 194, 45

\bibitem[{Steinhardt} \& {Elvis}(2010{\natexlab{a}})]{Steinhardt2010a}
{Steinhardt} C.~L., {Elvis} M., 2010{\natexlab{a}}, \mnras, 402, 2637

\bibitem[{Steinhardt} \& {Elvis}(2010{\natexlab{b}})]{Steinhardt2010c}
{Steinhardt} C.~L., {Elvis} M., 2010{\natexlab{b}}, \mnras, 406, L1

\bibitem[{Steinhardt} \& {Elvis}(2011)]{Steinhardt2010b}
{Steinhardt} C.~L., {Elvis} M., 2011, \mnras, 410, 201

\bibitem[{Steinhardt} \& {Silverman}(2011)]{Steinhardt2011a}
{Steinhardt} C.~L., {Silverman} J.~D., 2011, ArXiv:1109.0537

\bibitem[{Vestergaard} et~al.(2008){Vestergaard}, {Fan}, {Tremonti}, {Osmer} \&
  {Richards}]{Vestergaard2008}
{Vestergaard} M., {Fan} X., {Tremonti} C.~A., {Osmer} P.~S., {Richards} G.~T.,
  2008, \apjl, 674, L1

\bibitem[{Vestergaard} \& {Peterson}(2006)]{Vestergaard2006}
{Vestergaard} M., {Peterson} B.~M., 2006, \apj, 641, 689

\bibitem[{Wang} et~al.(2009){Wang}, {Dong}, {Wang} et~al.]{Wang2009}
{Wang} J., {Dong} X., {Wang} T., et~al., 2009, \apj, 707, 1334

\end{thebibliography}

\end{document}